\documentclass[12pt]{article}
\begin{document}
\hfill\vbox{\baselineskip14pt
            \hbox{\bf Feb. 2004}
            \hbox{\bf 022-P57}}
\baselineskip20pt
\vskip 0.2cm 
\begin{center}
{\Large\bf X-ray Absorption Near Edge Structure of FePt nanoclusters }
\end{center} 
\begin{center}
\large Sher~Alam$^{[1,2]}$\footnote{email:ALAM.Sher@nims.go.jp
:Tel:81-29-861-8650}
,~J. Ahmed$^{3}$,and ~Y. Matsui$^{1}$

\end{center}
\begin{center}
$^{1}${\it CSAAG, NIMS, Tsukuba, Ibaraki 305-0044, Japan}\\
$^{2}${\it Photonics, Nat.~Inst.~of~AIST, Tsukuba, Ibaraki 305-8568, Japan}\\
$^{3}${\it IMS, Univ. of Tsukuba, Tsukuba 305-8573, Japan}
\end{center}
\begin{center} 
\large Abstract
\end{center}
\begin{center}
\begin{minipage}{16cm}
\baselineskip=18pt
\noindent
X-ray Absorption Near Edge Structure [XANES] of FePt 
nanoclusters has been studied using a full multiple scattering,
self-consistent field [SCF], real-space Green`s function approach realized
via the powerful ab initio FEFF8 code. One purpose of our study is to
determine the sensitivity of Pt L3 edge with respect to the size and 
shape of the FePt nanoclusters. We also give the results
of the calculations with respect to the Fe L3 edge. Calculations
are made with and without core-hole for two main reasons, to
check and cross-check the FEFF code and also since in some cases 
it is known such as Pt clusters that better results are obtained 
without the core-hole. This is mainly because the screening 
electron will occupy empty d or f states and correspondingly reduce 
the white line intensity.
\end{minipage}
\end{center}
\vfill
\baselineskip=20pt
\normalsize
\newpage
\setcounter{page}{2}

\section{Introduction}
	In a previous note \cite{alam02} we have reported on the 
temperature dependent polarized X-ray absorption near-edge structure 
[XANES] spectra measured at the beamline BL13-B1 [Photon Factory, KEK],
in the fluorescence mode from 10K to 300 K for a very good quality
single crystal of LSCO co-doped with $1\%$ Zn at the copper site.
The main focus of \cite{alam02} was experimental, although some
initial results of the theoretical spectra generated by FEFF8.10 
\cite{ank98}, by using default settings [by intention] were given. 
In this note we continue our quest of using the FEFF code to study
systems of interest to us. In this short note we consider
small FePt clusters at both Pt and Fe L3 edges. The FePt clusters
are not only of academic interest due to their applications
to catalysis and useful magnetic properties i.e. for ultrahigh
density magnetic data storage media \cite{wu03}. The sensitivity
of Pt x-ray absorption near edge structure to small Pt clusters
has been recently given by \cite{ank03}.

	Thus the purpose of this note is to concentrate on the
theoretical calculation using the updated version of FEFF8, i.e.
FEFF8.20 \cite{ank02}. Here we report on our results of small
FePt clusters, giving the difference $\mu(E)$ spectra between
clusters and the appropriate bulk and between clusters at both
the Pt and Fe L3 edges. In addition to unpolarized spectra, we 
include the results taking polarization into account. 
Another aim is to provide a concrete example of the application 
of XANES and DOS calculations using FEFF. Indeed, although FEFF 
is a powerful code, its self-consistency 
must be demonstrated by applications to real systems systematically.
Moreover by actual calculations one can illustrate advantages
and disadvantages, and thus find ways of improving the code.

	The set-up of this paper is as follows, in the 
next section we outline some brief remarks about Electron
Multiple Scattering using FEFF.
Section three deals with results and discussion and finally
we state the conclusions. 

\section{Electron Multiple Scattering using FEFF}
In this note we concentrate on giving the difference $\mu(E)$
spectra. The value $\mu(E)$ is the main quantity in XANES.
Here we look at the basic definitions restricting ourselves
to brief comments. As is well-known
the primary quantities with which XANES calculations are concerned
are $\mu$,$\mu_{0}$, $\chi$, and $\rho_{li}(E)$\cite{ank98}
\begin{eqnarray}
\mu_{li}(E)=\mu_{li}^{0'}(E)[1+\chi_{li}^{'}(E)].
\label{f1}
\end{eqnarray}
We note that the prime is here for clarity since it
reminds us that it denotes final state quantities
in the {\em in the presence of screened hole}.
The central quantity in X-ray Absorption Spectroscopy [XAS] 
is the absorption coefficient $\mu(E)$. As is known there is a close and 
deep connection between XAS and {\em electronic structure}, which 
is indicated and implied by the resemblance of the contribution from a site, $i$ and
orbital angular momentum $l$ and the local $l$-projected
electronic density of states [LDOS] at site $i$
\begin{eqnarray}
\rho_{li}(E)=\rho_{li}^{0'}(E)[1+\chi_{li}(E)].
\label{f2}
\end{eqnarray}
However it is important to bear in mind that the
since core hole plays a significant role in calculation
of XAS, the similarity between XAS and LDOS cannot be 
regarded as an absolute.

	It is important to keep in mind that FEFF method starts
from the most fundamental quantity i.e. the Real Space Green's
Function and constructs the physical quantities of interest
from it \cite{ank98,ank02}. This is one of the code's main attraction 
since unlike band calculations it does not depend on symmetry. 
In this sense it is ideal for cluster physics\footnote{It is known 
that XANES and EXAFS signals are sensitive
to local structure. Indeed just as XRD is indicative of long-range
order, XANES and EXAFS carry signatures of short range order or
disorder.}. 


\section{Results and Discussion}
	Let us now give the results and analysis of our 
calculated theoretical results obtained with FEFF8.2, and the difference
spectra extracted with athena [version 0.8.024]\cite{new01}.
 For the background definitions and
formalism of FEFF8.2 we refer to its main reference \cite{ank02}.
We note that the Pt L3 edge roughly lies at 11,564 eV and 
the Fe L3 is located at approximately 706.8 eV. In this note
 we consider only the L3 edges of Pt and
Fe. We studied both nearly spherical [i.e.cuboctahedral] clusters 
and small non-spherical clusters with 3, 4, 5, and 8 atoms \cite{ank03}.  
The results of the nearly spherical clusters will not be discussed here.
Suffice is to say that we have used a spherical cluster of
55 atoms as a definition of bulk for Pt L3 edge, it has 1 central atom, 
12 atoms in first shell, 6 in the second shell, 24 in third shell,
and the remaining in the fourth shell. For the Fe L3 edge we have used
a 65 atoms clusters as defining the bulk. 
The 3 atoms planar cluster is written as (3,1) in the notation of Bigot
and Minot\cite{bm84}, for 4 atoms we can have the pyramid (4,1)
and planar rhombus configuration (4,2), the 5 atoms allows
again 2  possibilities, the square pyramid (5,2) and trigonal
bi-pyramid (5,4). Finally for the 8 atoms cluster we have three
configurations bi-capped octahedron (8,8), capped D5h bi-pyramid
(8,2,1) and trigonal prism with 2 square faces capped (8,2,4).

Fig.~\ref{fig1} displays the XANES difference spectra of the 
FePt 8 [xmu824] atom cluster with respect to the bulk PtFe at 
the Pt L3 edge as a function of the energy.
This data is for the unpolarized case. The spectrum in  Fig.\ref{fig1}
clearly shows that we can see the difference from the bulk at 
the L3 Pt edge, for the case of this FePt cluster. We have likewise 
calculated the difference spectra with respect to the bulk for other 
clusters and find significant deviations from the bulk.
Next we turn to the polarized case, see Fig.~\ref{fig2}. For illustration
the polarized result is given for another 8 atom cluster i.e. xmu821
cluster. The deviation
from the bulk is again significant. We note here that the polarized 
difference spectra do not vary significantly from the corresponding
unpolarized case, but the variation is significant between the polarized
and unpolarized case when the morphology is different, although the
number of atoms is the same, as is clear by comparing Figs.~\ref{fig1}
and ~\ref{fig2}.  

The results for the Fe L3 edge for both the unpolarized and the
polarized case are given in Figs.~\ref{fig3} and ~\ref{fig4}
respectively. 

Finally we consider the difference spectra between clusters
of same morphology and same number of atoms, but with different
atom species. To this end we take the difference spectra of
FePt cluster and a pure Pt cluster. The unpolarized case is
shown in Fig.~\ref{fig5} at the Pt L3 edge as a function of the 
energy, for the morphology (8,2,4). Likewise the polarized case
is indicated in  Fig.~\ref{fig6} for the morphology (8,8).

Several results are not shown here in this short presentation.
Interesting results are obtained if we take the difference
of polarized spectrum with respect to the corresponding unpolarized
case, in this instance one gets 
the ``polarization dichroic'' signal \cite{ank03}
which depends strongly on the shape of the cluster.
It is interesting to note that as intuitively expected the 
effect of polarization should be minimal for the nearly spherical 
clusters. We find this to be the case, thus providing a check on 
our calculations.

It is worth noting that the calculations were carried out with 
and without core-hole for two main reasons, to
check and cross-check the FEFF code and also since in some cases 
it is known such as Pt clusters that better results are obtained 
without the core-hole. This is since the screening 
electron occupies empty d or f states and correspondingly reduces 
the white line intensity.
  
\section{Conclusions}
The results and the analysis of the XANES of FePt nanoclusters using a full 
multiple scattering, self-consistent field SCF, real-space 
Green`s function approach realized via the powerful ab initio 
FEFF8 code have been outlined here. One purpose of our study was to
determine the sensitivity of Pt L3 edge with respect to the size and 
shape of the FePt nanoclusters, another was to give the results
of the calculations with respect to the Fe L3 edge. These points
have been demonstrated. Moreover results have also been given
when polarization is considered. In summary the XANES spectrum
shows sensitivity to the cluster size, morphology, and polarization
at the Pt and Fe L3 edge, for FePt clusters. 
\section*{Acknowledgements}
The work of Sher Alam is supported by the MONBUSHO
via the JSPS invitation program. We thank the developers
of the FEFF code J.~Rehr and company, plus Matt.~Newille
the Ifeffit developer and Bruce Ravel the developer
of Athena, Artemis and Atoms. J.~Rehr is thanked for providing
us their input files which were used in \cite{ank03} so that
crucial checks and cross-checks could be made.

\newpage
\begin{figure}
\caption{The difference xmu spectra of the (8,2,4) FePt cluster with respect
to the bulk PtFe at the Pt L3 edge, is plotted against energy.  
This unpolarized XANES data is generated with FEFF8.2.}
\label{fig1}
\end{figure}
\begin{figure}
\caption{The difference xmu spectra of the (8,2,1) FePt cluster
with respect to the corresponding bulk FePt at the 
Pt L3 edge. In this case polarization is included.}
\label{fig2}
\end{figure}
\begin{figure}
\caption{The difference xmu spectra of the (8,8) FePt cluster with respect
to the bulk FePt at the Fe L3 edge, is plotted against energy.}
\label{fig3}
\end{figure}
\begin{figure}
\caption{The difference xmu spectra of the (8,2,1) FePt cluster with respect
to the bulk FePt at the Fe L3 edge. Polarization is included.}
\label{fig4}
\end{figure}
\begin{figure}
\caption{The difference xmu spectra of the (8,2,4) FePt cluster with
respect to the (8,2,4) Pt cluster at the Pt L3 edge.}
\label{fig5}
\end{figure}
\begin{figure}
\caption{The difference xmu spectra of the (8,8) FePt cluster with
respect to the (8,8) Pt cluster at the Pt L3 edge when the 
polarization is taken into account.}
\label{fig6}
\end{figure}


\newpage
\begin{thebibliography}{99}
\bibitem{alam02}S.~Alam et~ al.,Physica~C~{\bf 378-381},
~(2002)~78-83.
\bibitem{wu03}A.~L.~Wu et al., App.~Phys.~Lett.~{\bf 82},
~(2003)~3475.
\bibitem{ank03}A.~L.~Ankudinov et al.,~J.~Chem.~Phys.~{\bf 116},
~(2003)~1911.
\bibitem{ank02}A.~L.~Ankudinov et al., Phys.~Rev.~{\bf B65},
~(2002)~104107.
\bibitem{ank98}A.~L.~Ankudinov et al., Phys.~Rev.~{\bf B58},
~(1998)~7565.
\bibitem{new01}M~.~Newille et al., J.~Synchrotron~Radiation{\bf 8},
~(2001)~322.
\bibitem{bm84}B~.~Bigot and C.~Minot, J.~Am.~Chem.~Soc.~{\bf 106},
~(1984)~6601.


\end{thebibliography}
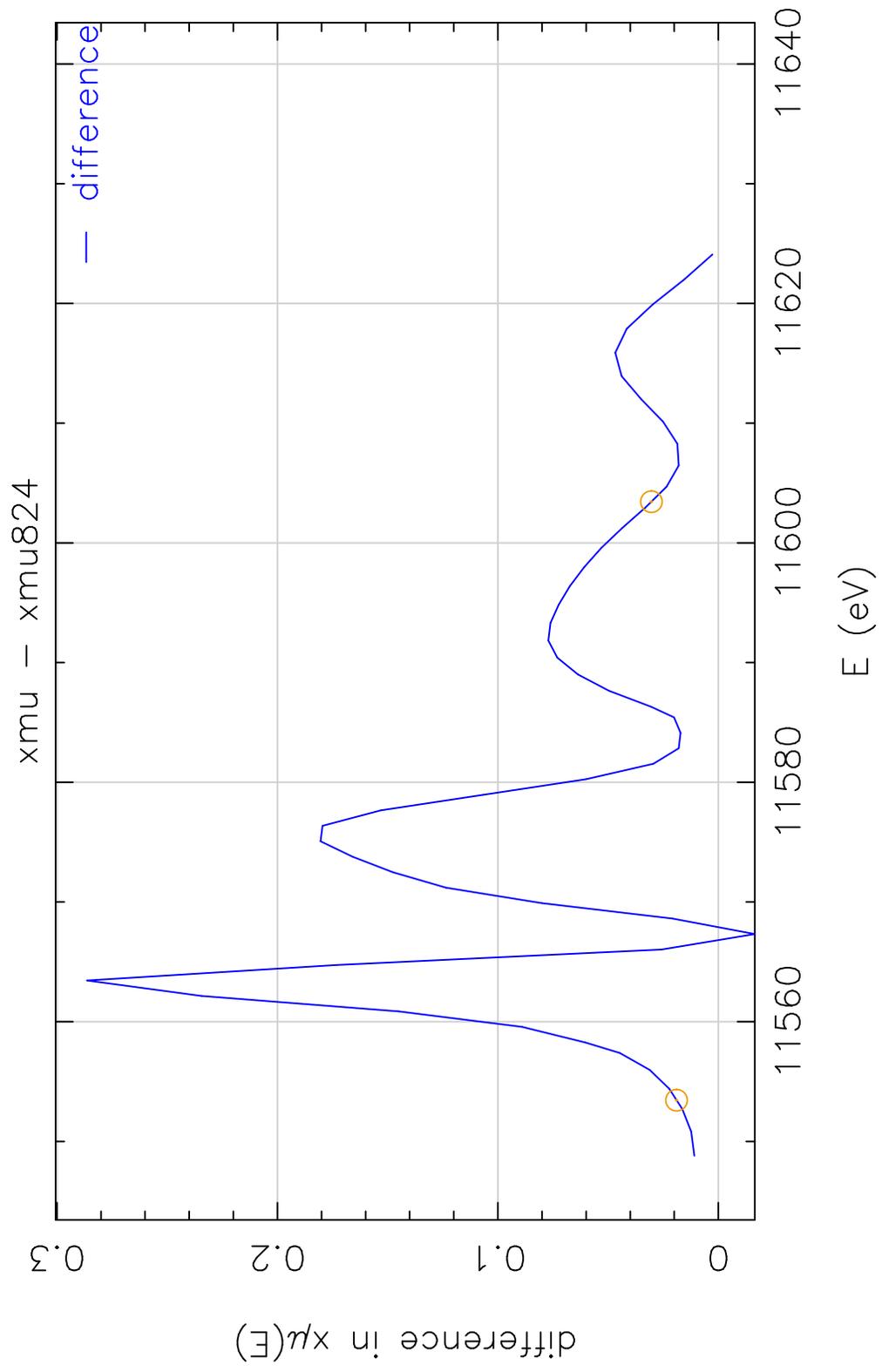
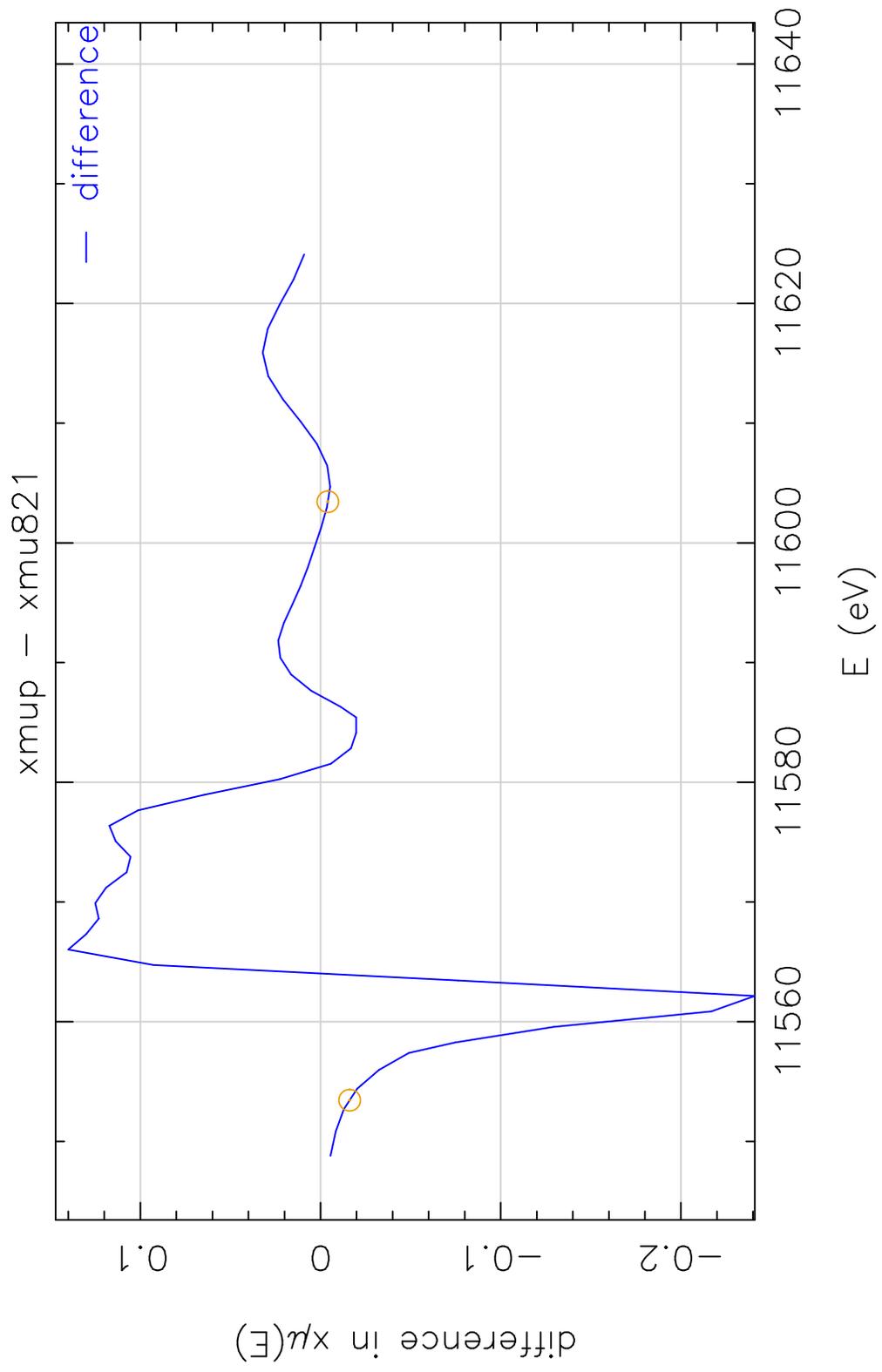
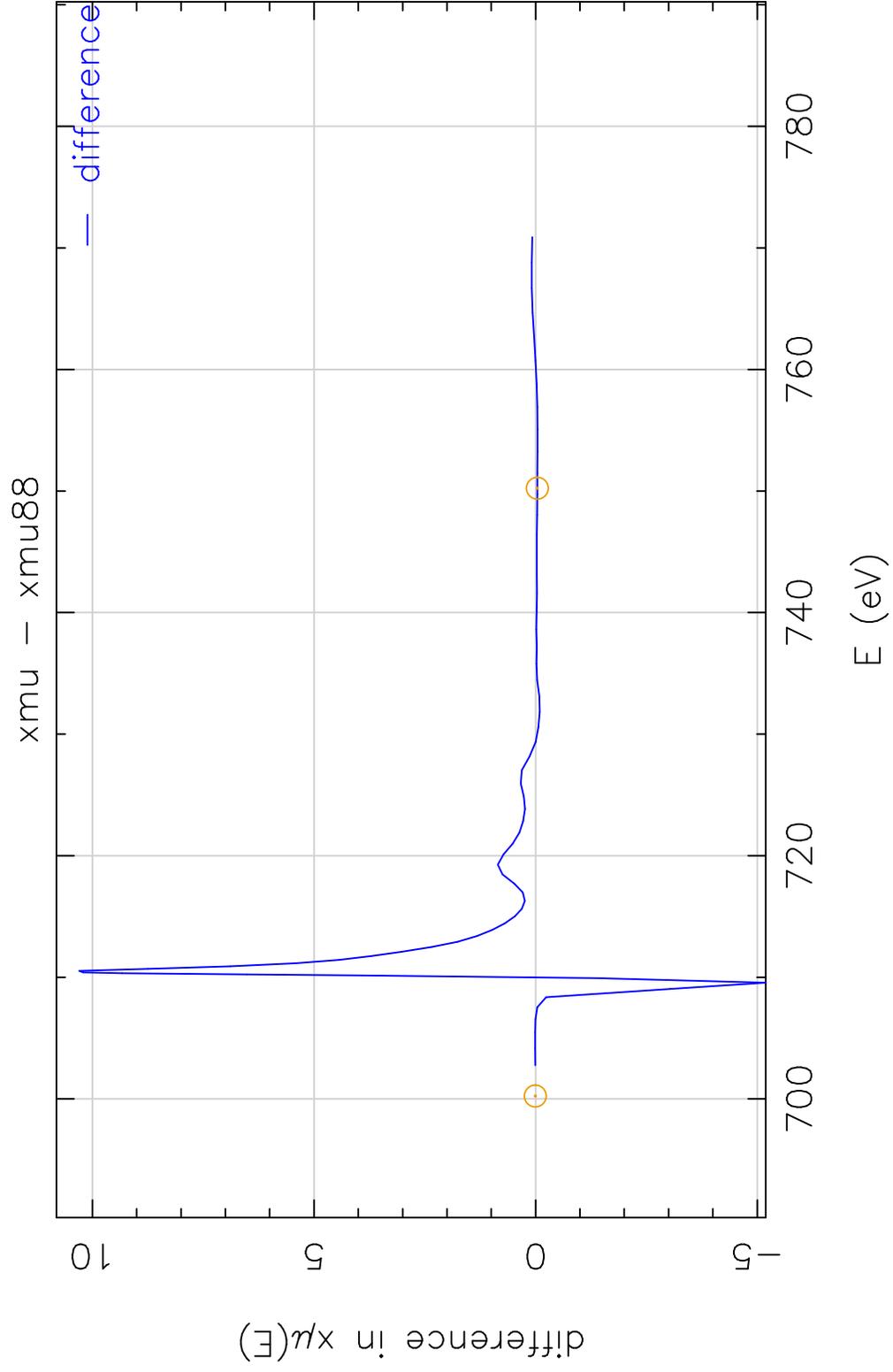
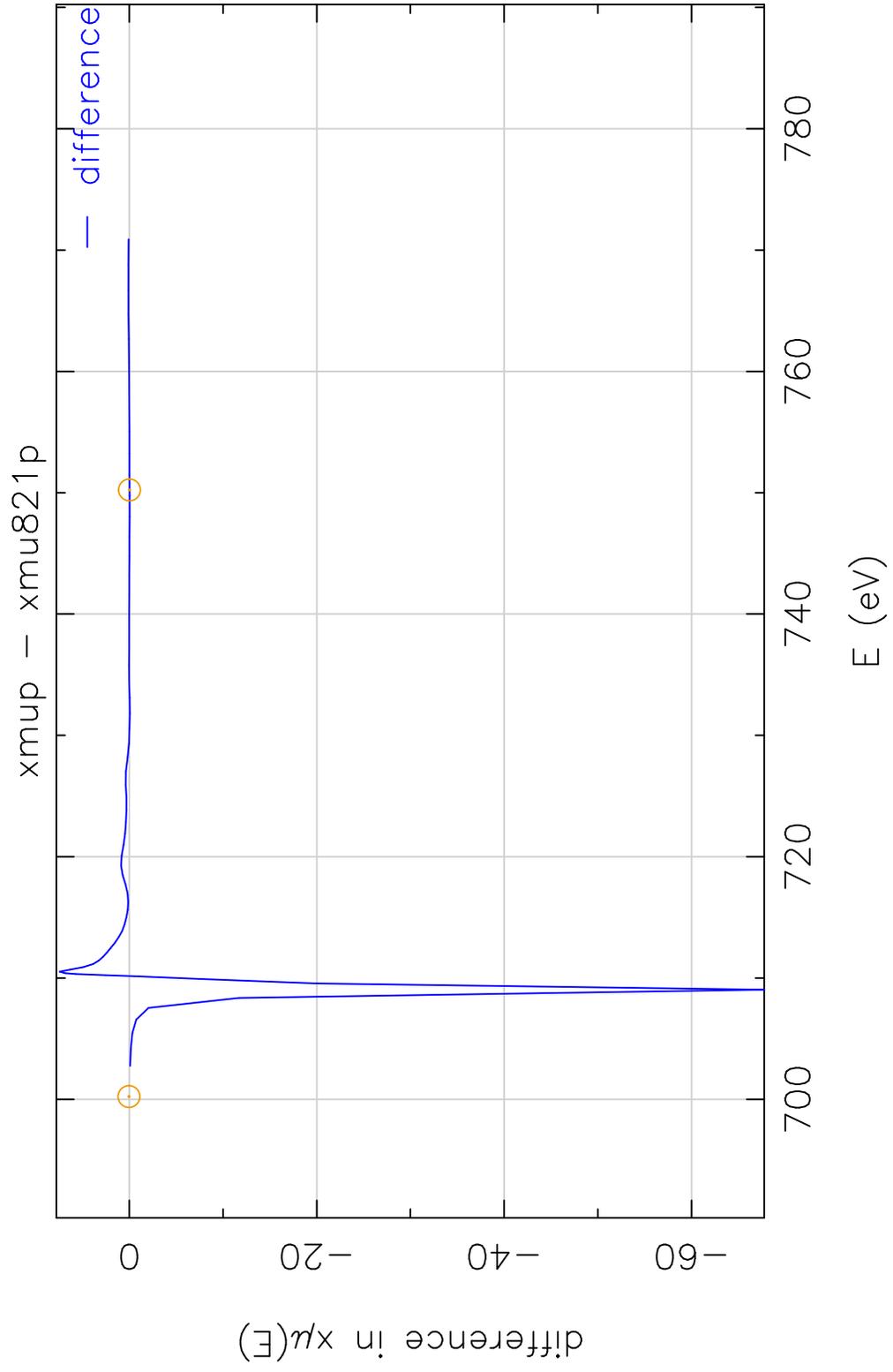
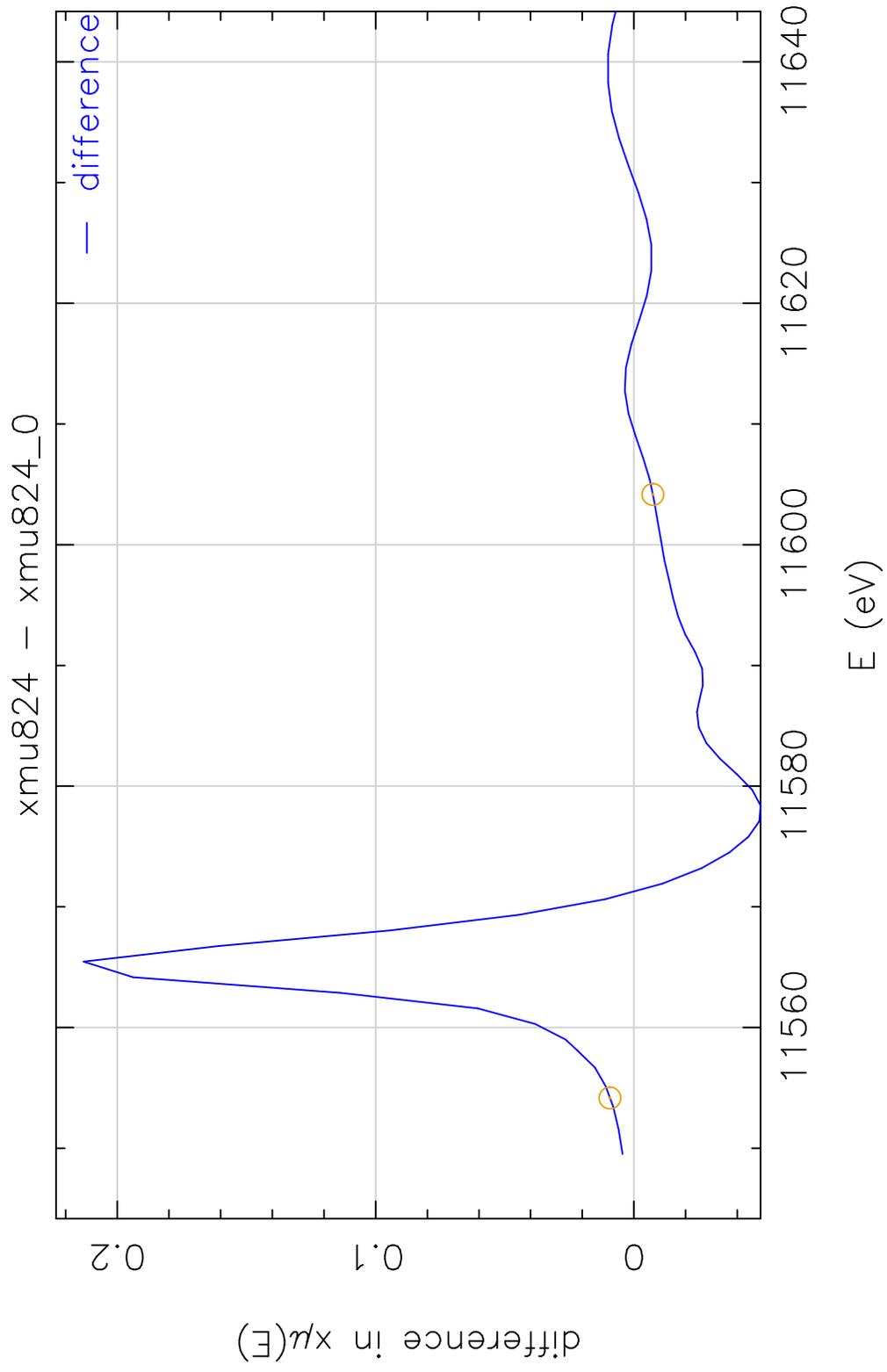
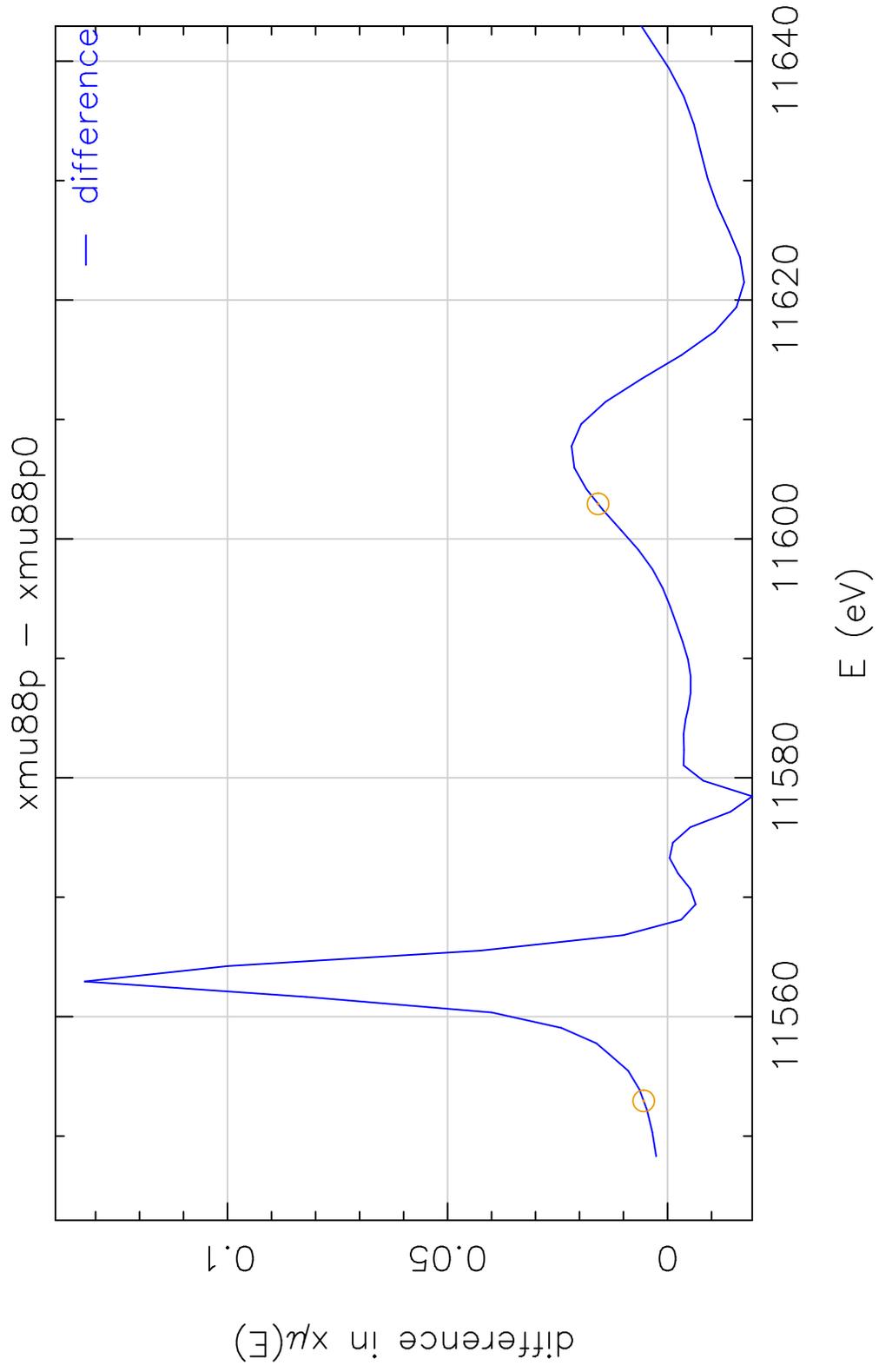
\end{document}